\documentclass[twocolumn]{aastex63}

\usepackage{color}
\usepackage{graphicx}
\usepackage{amsmath}
\usepackage{amssymb}
\usepackage{appendix}
\hypersetup{pdfauthor={Name}}

\def\ra#1#2#3{#1$^{\rm h}$#2$^{\rm m}$#3$^{\rm s}$}
\def\dec#1#2#3{$#1^\circ#2'#3''$}
\def\frb{FRB\,20201124A}

\def\oiii{{O\,{\sc iii}}}
\def\oii{{O\,{\sc ii}}}
\def\nii{{N\,{\sc ii}}}
\def\sii{{S\,{\sc ii}}}

\newcommand{\mdmunits}{{\rm pc \, cm^{-3}}} 
\newcommand{\dmunits}{$\mdmunits$}

\graphicspath{{./}{figures/}}

\shorttitle{FRB 20201124A Host Galaxy}
\shortauthors{Fong et al.}

\begin{document}
\sloppy

\title{Chronicling the Host Galaxy Properties of the Remarkable Repeating \frb}

\correspondingauthor{W. Fong}
\email{wfong@northwestern.edu}

\newcommand{\NU}{\affiliation{Center for Interdisciplinary Exploration and Research in Astrophysics (CIERA) and Department of Physics and Astronomy, Northwestern University, Evanston, IL 60208, USA}}

\newcommand{\Purdue}{\affiliation{Purdue University, 
Department of Physics and Astronomy, 525 Northwestern Avenue, West Lafayette, IN 47907, USA}}

\newcommand{\CfA}{\affiliation{Center for Astrophysics\:$|$\:Harvard \& Smithsonian, 60 Garden St. Cambridge, MA 02138, USA}}

\newcommand{\UCSC}{\affiliation{Department of Astronomy and Astrophysics, University of California, Santa Cruz, CA 95064, USA}}

\newcommand{\IS}{\affiliation{Centre for Astrophysics and Cosmology, Science Institute, University of Iceland, Dunhagi 5, 107 Reykjav\'ik, Iceland}}

\newcommand{\DAWN}{\affiliation{Cosmic Dawn Center (DAWN), Niels Bohr Institute, University of Copenhagen, Jagtvej 128, 2100 Copenhagen \O, Denmark}}

\newcommand{\PUCV}{\affiliation{Instituto de F\'isica, Pontificia Universidad Cat\'olica de Valpara\'iso, Casilla 4059, Valpara\'iso, Chile}}

\newcommand{\IPMU}{\affiliation{Kavli Institute for the Physics and Mathematics of the Universe (Kavli IPMU), 5-1-5 Kashiwanoha, Kashiwa, 277-8583, Japan}}

\newcommand{\PSU}{\affiliation{Department of Astronomy \& Astrophysics, The Pennsylvania State University, University Park, PA 16802, USA}}

\newcommand{\ICDS}{\affiliation{Institute for Computational \& Data Sciences, The Pennsylvania State University, University Park, PA, USA}}

\newcommand{\IGC}{\affiliation{Institute for Gravitation and the Cosmos, The Pennsylvania State University, University Park, PA 16802, USA}}

\newcommand{\Swin}{\affiliation{ Centre for Astrophysics and Supercomputing, Swinburne University of Technology, Hawthorn, VIC, 3122, Australia}}

\newcommand{\Curtin}{\affiliation{ International Centre for Radio Astronomy Research, Curtin University, Bentley, WA 6102, Australia}}

\newcommand{\MQ}{\affiliation{Department of Physics \& Astronomy, Macquarie University, NSW 2109, Australia}}

\newcommand{\MQAAAstro}{\affiliation{Macquarie University Research Centre for Astronomy, Astrophysics \& Astrophotonics, Sydney, NSW 2109, Australia}}

\newcommand{\CSIRO}{\affiliation{CSIRO, Space and Astronomy, PO Box 76, Epping NSW 1710 Australia}}
\author[0000-0002-7374-935X]{Wen-fai Fong}
\NU

\author[0000-0002-9363-8606]{Yuxin Dong}
\NU\Purdue

\author[0000-0001-6755-1315]{Joel Leja}
\PSU
\ICDS
\IGC

\author[0000-0003-3460-506X]{Shivani Bhandari}
\CSIRO

\author[0000-0002-8101-3027]{Cherie K. Day}
\Swin
\CSIRO

\author[0000-0001-9434-3837]{Adam T. Deller}
\Swin

\author[0000-0003-1913-3092]{Pravir Kumar}
\Swin

\author[0000-0002-7738-6875]{J.~Xavier~Prochaska}
\UCSC
\IPMU

\author[0000-0002-6895-4156]{Danica R. Scott}
\Curtin

\author[0000-0003-2149-0363]{Keith W. Bannister}
\CSIRO

\author[0000-0003-0307-9984]{Tarraneh Eftekhari}
\CfA

\author[0000-0002-5025-4645]{Alexa C. Gordon}
\NU

\author[0000-0002-9389-7413]{Kasper E. Heintz}
\IS\DAWN

\author{Clancy W. James}
\Curtin

\author[0000-0002-5740-7747]{Charles D. Kilpatrick}
\NU

\author[0000-0002-5053-2828]{Elizabeth K. Mahony}
\CSIRO

\author[0000-0003-3937-0618]{Alicia Rouco Escorial}
\NU

\author[0000-0003-4501-8100]{Stuart D. Ryder}
\MQ
\MQAAAstro

\author[0000-0002-7285-6348]{Ryan M. Shannon}
\Swin

\author[0000-0002-1883-4252]{Nicolas Tejos}
\PUCV

\received{2021 June 22}
\accepted{2021 August 31}
\submitjournal{Astrophysical Journal Letters}

\reportnum{astro-ph/2106.11993}

\begin{abstract}
We present the Australian Square Kilometre Array Pathfinder (ASKAP) localization and follow-up observations of the host galaxy of the repeating fast radio burst (FRB) source, \frb, the fifth such extragalactic repeating FRB with an identified host. From spectroscopic observations using the 6.5-m MMT Observatory, we derive a redshift of $z=0.0979 \pm 0.0001$, a star formation rate inferred from H$\alpha$ emission of SFR(H$\alpha$)~$\approx 2.1~M_{\odot}$~yr$^{-1}$, and a gas-phase metallicity of 12+log(O/H)$\approx 9.0$. By jointly modeling the 12-filter optical-mid-infrared (MIR) photometry and spectroscopy of the host, we infer a median stellar mass of $\approx 2 \times 10^{10}~M_{\odot}$, internal dust extinction of $A_V\approx 1-1.5$~mag, and a mass-weighted stellar population age of $\approx 5-6$~Gyr. Connecting these data to the radio and X-ray observations, we cannot reconcile the broad-band behavior with strong AGN activity and instead attribute the dominant source of persistent radio emission to star formation, likely originating from the circumnuclear region of the host. The modeling also indicates a hot dust component contributing to the MIR luminosity at a level of $\approx 10-30\%$. We model the host galaxy's star formation and mass assembly histories, finding that the host assembled $>90\%$ of its mass by 1~Gyr ago and exhibited a fairly constant SFR for most of its existence, with no clear evidence of past star-burst activity.
\end{abstract}


\section{Introduction}
\label{sec:intro}

Fast radio bursts (FRBs) are bright, millisecond-duration pulses detected almost exclusively at $\sim 0.1-8$~GHz frequencies \citep{lbm+07,tsb+13,cc19,phl19,CHIMECatalog21,gsp+18}. Perhaps their most enigmatic feature is that a fraction are observed to undergo repeat bursts from the same source (``repeaters''; \citealt{ssh+16}), while the vast majority have not been observed to repeat (apparent ``non-repeaters'' or ``one-off'' bursts; \citealt{smb+18}). It is not yet clear whether the collective population of FRBs originate from a single type of stellar progenitor or from multiple channels \citep{cc19,pwt+19}. However, population studies indicate that bursts from repeating FRBs are observed to have narrow bandwidths and longer temporal durations than those of apparent non-repeaters (\citealt{gsp+18,CHIME2019Repeaters,fab+20,Day2020,CHIMEMorphologies21}). Beyond these attributes, scant clues exist for differences in their physical origins. 

Alongside studies of their emission properties, examining the environments of FRBs on sub-parcsec to kiloparsec scales can be equally informative. Thus far, only a fraction of known extragalactic repeating FRBs have been localized to host galaxies \citep{clw+17,tbc+17,mpm+20,mnh+20,hps+20,bgk+21,bha+21,lna+21,bkm+21}\footnote{FRB\,20201124A is the fifth announced repeating FRB with a host galaxy. In total, there are eight such FRB sources known as of 2021 August.}. Much closer by, FRB-like emission has been detected from a Milky Way magnetar, SGR 1935+2154 \citep{CHIME-FRB200428,brb+20}. All identified repeating FRB hosts have evidence for low to modest ongoing star formation rates of $\approx 0.06-2\,M_{\odot}$~yr$^{-1}$ \citep{gpm+04,bsp+20,hps+20}, several exhibit spiral arm morphologies \citep{mfs+20,bgk+21,tgk+21}, and their stellar populations span a range of stellar masses, $\approx 10^{8}-10^{10.5}\,M_{\odot}$ \citep{dwb+08,bsp+20,hps+20,mfs+20}. At face value, these characteristics, coupled with the absence of any quiescent host galaxy identifications for repeating FRBs, may indicate that FRBs are connected to host galaxies with ongoing star formation. However, studies of their more local environments reveal a rich diversity. For instance, the discovery of the repeating FRB\,20200120E in an old (9.1~Gyr) globular cluster on the outskirts of the grand design spiral galaxy M81 \citep{bgk+21,kmn+21} and the detection of the repeating FRB\,20121102A embedded in a star-forming knot in its dwarf host galaxy \citep{bta+17} seemingly represent polar opposite local environments. If all repeaters discovered to date originate from the same type of progenitor, then the progenitor model must accommodate the observed diversity of both local and galactic environments. Additionally, any connection in progenitors to the population of apparent non-repeaters remains opaque.

The number of repeating FRBs with identified hosts are still few in number. We are thus motivated to characterize the environments of any new repeating FRBs to understand the full spectrum of environments that give rise to these events. \frb\ was first discovered by the Canadian Hydrogen Intensity Mapping Experiment FRB (CHIME/FRB) collaboration \citep{atel14497,TNSdiscovery} on 2020 November 24 UTC at 08:50:41. On 2021 March 31, the CHIME/FRB collaboration reported that the \frb\ source had repeated and was entering a period of high activity \citep{atel14497}.  Numerous radio facilities have since reported the detection of repeating bursts from the same source, including ASKAP \citep{atel14502,atel14508}, the Five-hundred-meter Aperture Spherical radio Telescope (FAST; \citealt{atel14518}), the Very Large Array (VLA; \citealt{atel14526}), the Upgraded Giant Metrewave Radio Telescope (uGMRT; \citealt{atel14538,mbm+21}), the Stockert 25-m Radio Telescope \citep{atel14556}, Onsala Space Observatory \citep{atel14605}, and the Allen Telescope Array \citep{atel14676}. To date, over 1700 bursts from the \frb\ source have been reported between its discovery in 2020 November 24 to 2021 August 10\footnote{Includes data from ATELs, \citet{mbm+21}, \citet{pbt+21} and the CHIME repeater catalog: \url{https://www.chime-frb.ca/repeaters/FRB20201124A}}. In addition, uGMRT and VLA observations uncovered a persistent radio source (PRS) at 650~MHz, 3~GHz and 9~GHz \citep{atel14529,atel14549,rll+21} reported to be unresolved on arcsecond scales.

Here we report on the arcsecond-precision localization based on three bursts from the actively repeating \frb\ source detected by the Australian Square Kilometre Array Pathfinder (ASKAP) telescope as part of the Commensal Real-time ASKAP Fast Transients (CRAFT) project \cite[][]{macquart+10}. In Section~\ref{sec:obs}, we describe the community observations to date and introduce the properties and localization of three ASKAP bursts, as well as their joint localization. Here, we also describe the observations and follow-up observations of the host galaxy. In Section~\ref{sec:host}, we model the properties of the host galaxy, construct its star formation and mass assembly histories, and explore the origin of the persistent radio emission by connecting the broad-band observations. In Section~\ref{sec:disc}, we discuss the host of \frb\ in the context of other repeating FRBs environments, and implications for the progenitors. We summarize in Section~\ref{sec:conc}. We note that \citet{rll+21} and \citet{pbt+21} present independent analyses of the host galaxy and persistent radio emission of this FRB, and our results are broadly consistent with those works.

Unless otherwise stated, all observations are reported in the AB magnitude system and have been corrected for Galactic extinction in the direction of the FRB using $A_{\rm V}=1.95$~mag \citep{fm07} and the \citet{ccm89} extinction
law.  We employ a standard cosmology of $H_{0}$ = 69.6~km~s$^{-1}$~Mpc$^{-1}$, $\Omega_{M}$ = 0.286 \citep{blw+14}, and $\Omega_{\lambda}$ = 1-$\Omega_{M}=0.714$.

\begin{figure*}[t]
\centering
\includegraphics[width=1.0\textwidth]{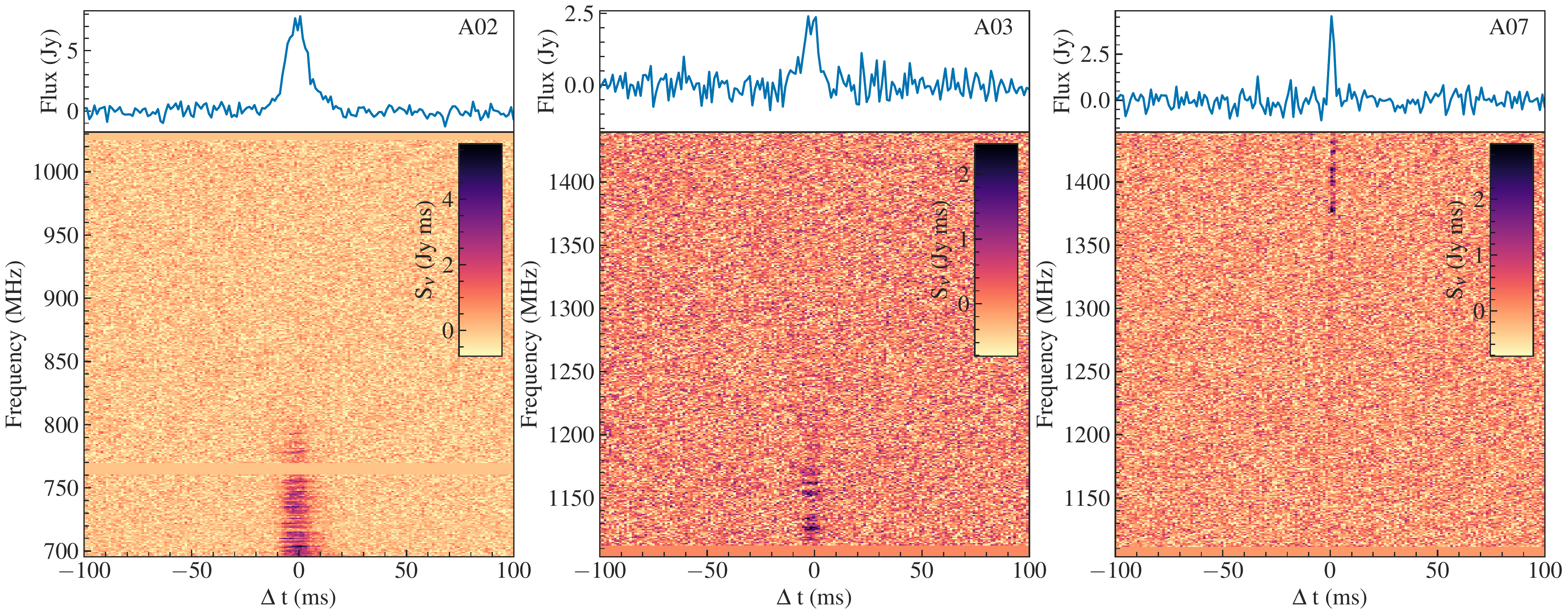}
\caption{Dynamic spectra of three bursts detected by ASKAP from the \frb\ source (A02, A03, and A07, from left to right). All bursts have been de-dispersed to their respective optimized DMs. In each sub-plot, the bottom panel shows the dynamic spectrum and the top panel shows the frequency-averaged pulse profile.
\label{fig:burst_waterfall}}
\end{figure*}

\begin{deluxetable*}{ccccccccccccc}[!t]
\tablecaption{Properties of Three ASKAP Bursts from the \frb\ Source \label{tab:loc}}
\tablecolumns{10}
\tablewidth{0pt}
\tablehead{
\colhead{Date} &
\colhead{TNS Name} & 
\colhead{Burst No.} & 
\colhead{Frequency} & 
\colhead{Fluence} & 
\colhead{DM} & 
\colhead{RA} & 
\colhead{Decl.} &
\colhead{$\sigma_{\rm maj}$} &
\colhead{$\sigma_{\rm min}$} &
\colhead{$\sigma_{\rm PA}$}  \\
\colhead{(UT)} & 
\colhead{} &
\colhead{} &
\colhead{(MHz)} & 
\colhead{(Jy~ms)} & 
\colhead{(pc~cm$^{-3}$)} & 
\colhead{(J2000)} & 
\colhead{(J2000)} & 
\colhead{($''$)} & 
\colhead{($''$)} & 
\colhead{(deg)} 
}
\startdata
2021-04-01 & 20210401A & A02 & 864.5 & $187 \pm 12$  & $412 \pm 3$ & \ra{05}{08}{03.48} & $+$\dec{26}{03}{38.4} & 1.3 & 0.7 & 140 \\
2021-04-02 & 20210402A & A03 & 1271.5 & $22 \pm 3$ & $414 \pm 3$ & \ra{05}{08}{03.67} & $+$\dec{26}{03}{39.5} & 1.6 & 0.9 & 42 \\
2021-04-04 & 20210404B & A07 & 1271.5 & $11 \pm 2$ & $414 \pm 3$ & \ra{05}{08}{03.55} & $+$\dec{26}{03}{39.1} & 1.1 & 0.7 & 13 \\
Combined  &  & & & & & \ra{05}{08}{03.54} & $+$\dec{26}{03}{38.4} & 0.9 & 0.8 & 171   \\
\hline
\enddata
\tablecomments{Uncertainties correspond to $1\sigma$ confidence. $\sigma_{\rm maj}$ and $\sigma_{\rm min}$ are the uncertainties along the semi-major and semi-minor axes respectively, and then rotated by the $\sigma_{\rm PA}$, defined as the position angle East of North.}
\end{deluxetable*}

\section{Observations and Burst Properties}
\label{sec:obs}

\subsection{ASKAP Burst Detections and Localization}

\begin{figure*}[t]
\centering
\includegraphics[width=0.7\textwidth]{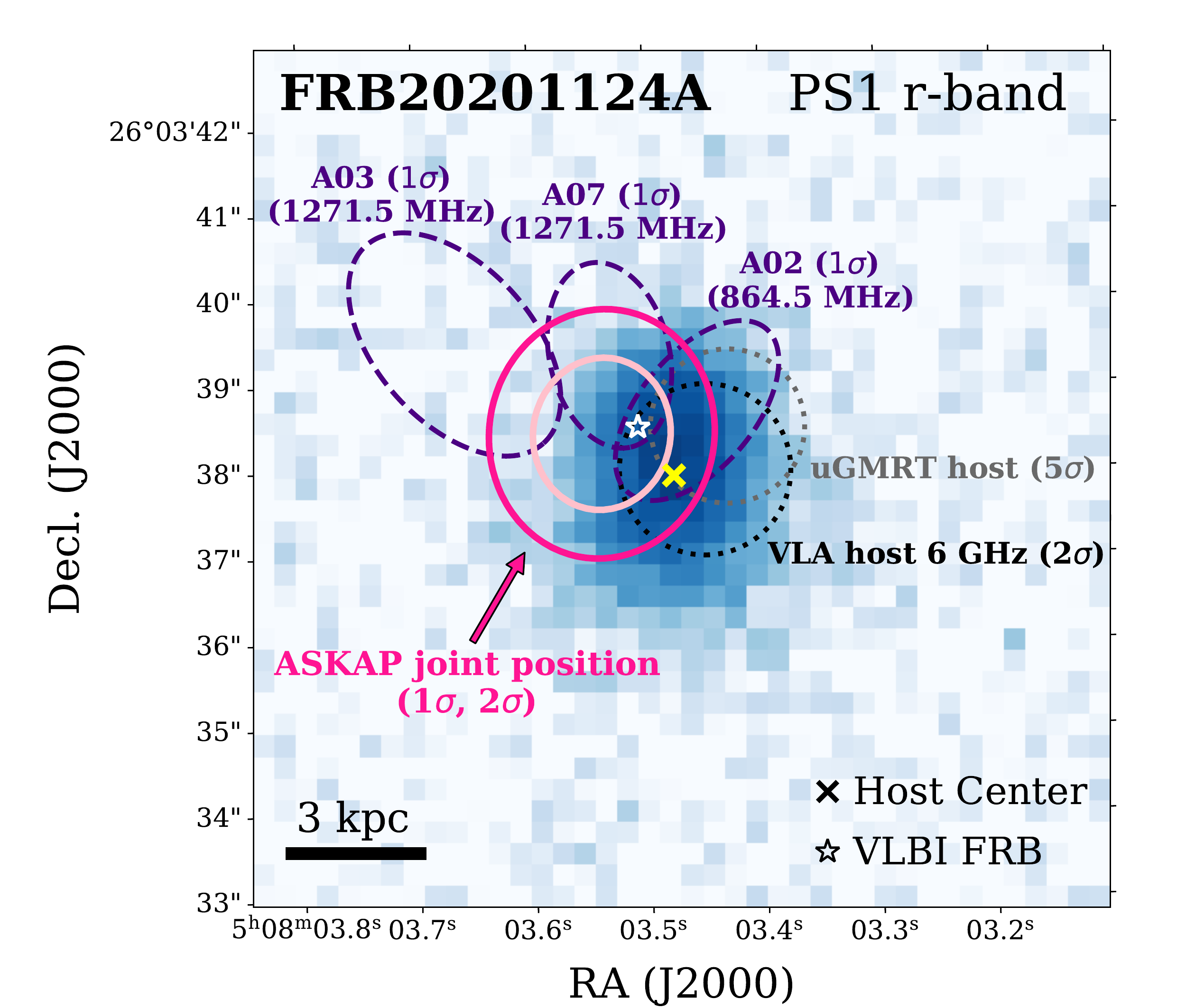}
\vspace{-0.1in}
\caption{Archival Pan-STARRS1 $r$-band imaging of the putative host galaxy of FRB\,20201124A. The positions of the three bursts from ASKAP are plotted (purple dashed lines; $1\sigma$), along with the joint probability map (solid pink contours; $1\sigma$ and $2\sigma$, respectively). The ASKAP joint position is consistent within the $1\sigma$ contour with the available VLBI position (white star), and $\approx 1.3$~kpc offset from the measured center of the host galaxy (yellow `x'). Dotted lines denote the persistent radio emission positions reported by the VLA ($2\sigma$; total error; \citealt{rll+21}) and uGMRT ($5\sigma$; statistical uncertainty; \citealt{atel14529}), which are nominally consistent with the host galaxy center and also consistent with the joint FRB position.
\label{fig:host_image}}
\end{figure*}

We have been conducting surveys for FRBs with the ASKAP as part of the CRAFT project.  ASKAP is a 36-antenna array, with each antenna equipped with a phased-array receiver capable of forming 36 beams on the sky for a total field of view of $\approx 30$\,deg${^2}$. FRB searches are currently conducted on the incoherent sum of  total intensities for each beam from each antenna. The searches reported here were conducted on data with $1.2$~ms time resolution, $1$~MHz spectral resolution, and a total bandwidth of $336$~MHz. The data were recorded at a central frequency of either $864.5$~MHz (low-band) or $1271.5$~MHz (mid-band) on the incoherent sum of $23$ or $24$ antennas. 
The searches were initially conducted with the central ASKAP beam pointed at the best-fit position of the initial CHIME localization\footnote{Searches were conducted with both hexagonal close pack and square beam arrangements.  At $864.5$~MHz, the beams were separated by $1.05$~deg.  At $1271.5$~MHz, the beams were separated by 0.9~deg.} \cite[][]{atel14497}.
The ASKAP interferometric position for the burst, which differed by $\sim$8 arcminutes from the initial pointing, was used for observations of subsequent bursts after it had been measured. 

The searches were conducted in real-time using a graphical processing unit-accelerated implementation \cite[][]{bzq+19} of the fast dispersion measure transform \cite[][]{zo17}, optimized for low latency (sub-second) detections. 
Further details of the search methods can be found in \cite{bdp+19}.
Candidates identified by the pipeline trigger a download of the $3.1$\,s-duration voltage buffer for the candidate beam from each antenna. These voltages are correlated, calibrated, and imaged to interferometrically determine the position of the burst, following the procedure in \cite{Day2020}.    

Between 2021 April 1 and April 7 UTC, ASKAP detected 11 bursts from the \frb\ source.
For five of the bursts, real-time triggers were generated that led to the download of the voltage buffers that are required for localization.\footnote{Of the other six bursts, several were identified in real time but failed a temporal width test that prevents excessive download triggers caused by Radio Frequency Interference, while others were sub-threshold and were only identified in {\em post-facto} analysis.} From the five real-time triggers, two downloads experienced a partial networking failure which led to a reduced subset of antenna voltages being acquired (7 and 11 respectively).  While the FRB itself could still be detected in the resultant images, the reduced sensitivity meant that the background continuum radio sources used to refine the absolute astrometric calibration of the FRB image (described further below) could not be detected with sufficiently high signal-to-noise. This leaves three bursts suitable for localization.
The results here are derived from the second, third, and seventh bursts detected by ASKAP (A02, A03, and A07), while the full set of burst properties will be described in an upcoming work. 
We previously reported on the localizations of the two bursts detected by ASKAP on 2021 April 1 and 2 UTC (FRB\,20210401A: A02 and FRB\,20210402A: A03, respectively; \citealt{atel14515,atel14592}).  Here, we also introduce the discovery and localization of a third burst detected by ASKAP on 2021 April 4 UTC (FRB\,20210404B: A07). A02 was detected in the low-band at $864.5$~MHz with a fluence of $187 \pm 12$~Jy~ms, while the latter two bursts were detected in the mid-band at $1271.5$~MHz, with significantly lower fluences of $22 \pm 3$~Jy~ms and $11 \pm 2$~Jy~ms, respectively. The dispersion measure (DM) is DM=$412 \pm 3$~pc~cm$^{-3}$ (A02), and DM=$414 \pm 3$~pc~cm$^{-3}$ (A03 and A07), in agreement with the initial reported value from CHIME/FRB of DM=$413.52 \pm 0.05$~pc~cm$^{-3}$ \citep{atel14497}. The burst dynamic spectra and frequency-averaged pulse profiles for the three bursts are displayed in Figure~\ref{fig:burst_waterfall}, and their properties are listed in Table~\ref{tab:loc}.
The spectral, temporal, and polarimetric  properties of the bursts will be discussed in future papers. 

 While the FRB was strongly detected in every image, the absolute positional registration of the FRB images was hampered by two issues: the low observing elevation at the time of A02 and A03 ($\sim$20 degrees, which leads to a larger and much more elongated synthesized beam) and the paucity of bright background sources used to estimate and correct for any systematic position shift in the FRB image.  The latter issue was particularly problematic in the mid band (A03 and A07), due to the negative spectral index of typical background sources.

To account for the highly elongated synthesized beams, for each burst, we utilized a coordinate frame aligned with the position angle of the synthesized beam when estimating the systematic positional shift via a weighted mean of the positional offsets seen in the background radio sources.  As each source utilized from the ASKAP image is consistent with being unresolved (any clearly resolved sources are rejected, as the position centroid at ASKAP angular resolution/frequency may not match the position centroid used to calculate the catalog value) and hence has a position angle close to that of the synthesized beam, this approach leads to a minimal correlation in the offsets in the two coordinates.  \citet{Day21} show that a simple weighted mean of the offsets obtained using background radio sources overestimates the accuracy to which the mean image shift can be determined when using images that are typical for ASKAP FRB detections, and determine that a scaling factor of 1.79 corrects for this on average.  For A03 and A07, at elevations of $\sim$20 degrees, however, we find that the uncertainty is likely still underestimated after the application of this scaling factor (based on the reduced $\chi^2$ of the weighted mean of the offsets).  An increased scale factor of 3 (rather than the global average of 1.79) yields a reduced $\chi^2$ that is consistent with expectations for each individual FRB field, and so we conservatively use that for the systematic uncertainties that dominate the overall positional uncertainty shown in Table~\ref{tab:loc} and Figure~\ref{fig:host_image}.

We then place an improved constraint on the source position by evaluating the likelihood on a positional grid, using the probability from all three individual ASKAP localizations at each grid point.  This results in a best-fit ASKAP position for the FRB of RA$=$\ra{05}{08}{03.54(5)} and Dec$=+$\dec{26}{03}{38.4(9)}, where the best-fit $\chi^2$ was 3.4 for 4 degrees of freedom and the uncertainties reflect 68\% confidence intervals. 
The joint ASKAP localization is consistent with the reported VLBI position \citep{atel14603} within the 68\% confidence contour, and the $\chi^2$ indicates that despite the localization of burst A03 differing from the VLBI position by $\sim$1.5$\sigma$ in one coordinate, the uncertainties are overall well estimated. The positions and burst properties are listed in Table~\ref{tab:loc}, and the positions are plotted in Figure~\ref{fig:host_image}.

\begin{figure*}[!t]
\centering
\includegraphics[width=\textwidth]{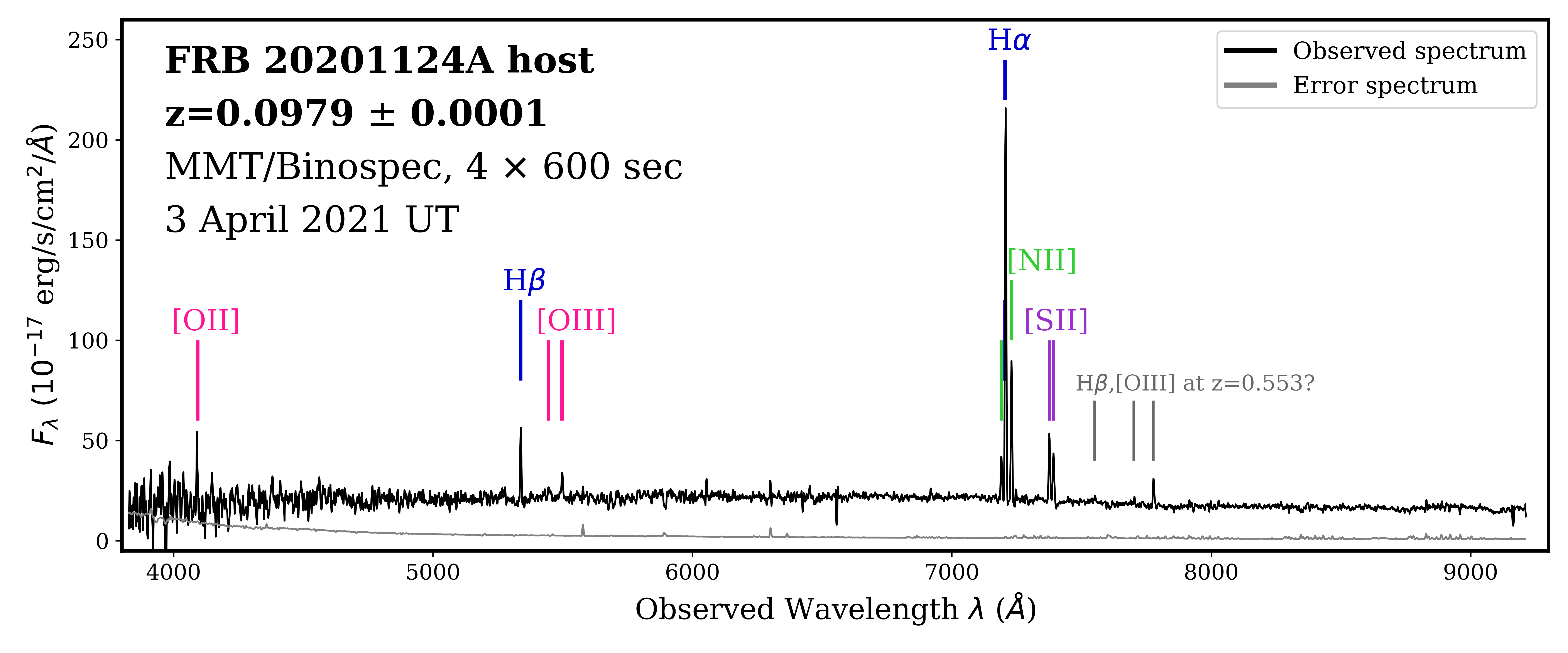}
\vspace{-0.3in}
\caption{MMT/Binospec spectroscopy of the putative host galaxy of \frb. The locations of several emission lines are marked, at a common redshift of $z=0.0979 \pm 0.0001$. We also detect a few emission features redward of $7500$\AA\ (most prominently at $7778$\AA), which we tentatively identify as H$\beta$ and [\oiii] emission from a background, spatially coincident but otherwise unrelated galaxy at $z=0.5531$.
\label{fig:host_spec}}
\end{figure*}

\subsection{Host Galaxy Association and Redshift Determination}
\label{sec:redshift}

The combined ASKAP position from \frb\ is spatially consistent with the galaxy SDSS~J050803.48+260338.0 (first reported in \citealt{atel14515}) in the Sloan Digital Sky Survey (SDSS) catalog with tabulated photometric redshifts ($z_{\rm phot}$) of $0.11 \pm 0.03$ and $0.08 \pm 0.02$ from the Panoramic Survey Telescope and Rapid Response System Data Release 1 (Pan-STARRS DR1; \citealt{cmm+16,bsf+19}) and SDSS Data Release 16 catalog \citep{SDSS2020}, respectively. Figure~\ref{fig:host_image} shows an archival Pan-STARRS $r$-band image, along with the individual and joint ASKAP burst positions, the VLBI burst position, and the uGMRT and VLA PRS positions.

We use \texttt{Source Extractor} \citep{SExtractor} on the Pan-STARRS $r$-band image to derive a host galaxy centroid of RA=\ra{05}{08}{03.477}, Dec=$+$\dec{26}{03}{37.93} with an uncertainty of $\approx 11-13$~mas in each coordinate. The best-fit projected angular offset between the centroid of the combined ASKAP position and host center is $\approx 0.97''$, but the relatively low ASKAP localization precision for this source means that the FRB position cannot be excluded from consistency with the host centroid at a 95\% confidence level. The VLBI position \citep{atel14603}, however, affords a highly-precise measured offset from the host center of $0.71 \pm 0.03''$, taking into account the reported VLBI uncertainty, the host centroid uncertainty, and the median astrometric tie uncertainty of Pan-STARRS to Gaia of 22~mas \citep{msf+20}. Using the angular separation and the extinction-corrected magnitude, $r \approx 17.84$~mag (Table~\ref{tab:hostphot}), we calculate a low probability of chance coincidence, $P_{\rm cc} \approx 6.9 \times 10^{-5}$ \citep{bkd02} pointing to a robust association\footnote{We use Equation~1 in \citet{ber10} with $m=17.84$~mag and $\delta R=0.71''$}. This association is strengthened by adopting the Bayesian formalism PATH (Probabilistic Association of Transients to their Hosts; \citealt{PATH})
and using the PS1 catalog to derive a posterior probability of association, $P(O|x)\approx 0.997$ (where $P(O|x)>0.95$ may be considered a secure association), assuming a 10\% prior probability that the host galaxy is not in the list of catalogued galaxies. For the PATH probability, we assume an ``inverse'' prior in which brighter candidate hosts have higher prior probability according to their number density on the sky (Equations 4 and 12 in \citealt{PATH}) and assumed an underlying ``exponential'' distribution model for the offsets (Equation 14 in \citealt{PATH}). We thus confidently assign SDSS~J050803.48+260338.0 as
the host galaxy of \frb.

We initiated spectroscopic observations of the host of \frb\ with the Binospec imaging spectrograph \citep{Binospec} mounted on the 6.5-m MMT Observatory atop Mount Hopkins, Arizona (Program UAO-G195-21A; PI: Fong) on 2021 April 3 UTC. The preliminary results of these observations were reported in \citet{atel14516}. We obtained $4 \times 600$~s of exposures with the 270l grating and the LP3800 spectroscopic filter to mitigate the effects of second-order light contamination. The central wavelength was 6500~\AA\ to cover a wavelength range of $\approx 3800-9200$~\AA, and the $1''$ slit was oriented at a position angle 145$^{\circ}$ East of North. 
We used the Python Spectroscopic Data Reduction Pipeline (PypeIt; \citealt{PypeIt}) for data processing on the overscan-subtracted images from the MMT archive. In PypeIt, we apply a flat-field correction and perform wavelength calibration and spectral extraction (using the {\tt boxcar} method with $2.4\arcsec$ radius, in order to include all the extended line emission flux). 
We applied an absolute flux calibration using observations of the spectrophotometric standard Feige~34 taken on the same night. We co-add the 1D spectra and scaled the flux of the resulting co-addition to match the galaxy's measured $r$-band magnitude. Finally, we apply a telluric correction using an atmospheric model. We correct the spectrum for Galactic extinction using the \citet{ccm89} extinction law and $A_V=2.024$~mag in the direction of the burst. The final 1D spectrum of the host of \frb\ is displayed in Figure~\ref{fig:host_spec}.

We detect several clear emission features: [\oii]$\lambda3727$, H$\beta\lambda4861$, [\oiii]$\lambda5007$, H$\alpha\lambda6563$, [\nii]$\lambda6548$ and $\lambda6583$, and [\sii]$\lambda6716$ and $\lambda6731$. From these lines, we derive a common redshift of $z=0.0979 \pm 0.0001$, consistent with the SDSS and PS1 photometric redshifts. At this redshift, the projected physical offset between the VLBI burst position and host center is $1.29 \pm 0.05$~kpc. Using a value of DM$_{\rm FRB} \approx 414$\,\dmunits, the DM for the Galactic ISM in the direction of the FRB of ${\rm DM}_{\rm ISM} = 123$\,\dmunits\  
\citep{ne2001}, and ${\rm DM}_{\rm MW, Halo}=50$\,\dmunits\ for the Milky Way halo \citep{pz19} and assuming a DM for the host of DM$_{\rm host}=50(1+z)$\,\dmunits, we find that the cosmic DM is DM$_{\rm cosmic} \approx 186$\,\dmunits. This derived value of DM$_{\rm cosmic}$ is in excess of the Macquart relation (which provides a relationship between DM$_{\rm cosmic}$ and $z$; \citealt{mpm+20}), which predicts DM$_{\rm cosmic} \approx 82$\,\dmunits. This can be reconciled if the DM intrinsic to the host or FRB environment is significantly larger than the value for DM$_{\rm host}$ assumed here, e.g., DM$_{\rm host}=100(1+z)$\,\dmunits.

We also remark on the presence of faint emission features redward of $7500$~\AA, with the strongest feature at $7777.78$~\AA\ (Figure~\ref{fig:host_spec}). These features do not correspond to any known emission lines at $z=0.0979$. Instead, we tentatively identify these features as H$\beta$ and [\oiii]$\lambda 4959$ and $\lambda5007$ from a {\it background} (likely unrelated, see below) galaxy at a common redshift of $z=0.5531$\footnote{We note that given the high Galactic extinction, and the low S/N with which we detect the redder emission features, the lack of detected [OII] at the same redshift is not surprising.}. The strongest of the features is visible in each of the individual exposures and has a spatial offset of $\approx 2$~pixels from the host galaxy continuum in all frames. Using the known position angle of the slit, we determine that the emission originates from a source $\approx 0.5''$ offset to the northwest of the host galaxy center in projection. This emission is not consistent with the VLBI position of the FRB, and is not readily visible in any of the available archival imaging. 
Adopting the Macquart relation and an estimate of
its intrinsic scatter \citep[parameterized with $F=0.3$;][]{mpm+20},
the PDF for the cosmic DM $P({\rm DM_{cosmic}}|z)$
yields a lower bound of 
${\rm DM}_{\rm cosmic}^{\rm min} = 311$~\dmunits\
(95\%\ c.l.\ for $z=0.55$).
Combined with ${\rm DM}_{\rm ISM} = 123$\,\dmunits \citep{ne2001},
we recover DM$_{\rm FRB} > 434$\,\dmunits, which
exceeds the measured value even before 
including the Galactic halo or any
host DM contribution.
Therefore, we conclude this putative galaxy is not the host of \frb.

\begin{deluxetable*}{cccccc}[t!]
\linespread{1.1}
\tablecaption{\frb\ Host Galaxy Photometry \label{tab:hostphot}}
\tablecolumns{6}
\tablewidth{0pt}
\tablehead{
\colhead{Source} &
\colhead{Filter/Band} &
\colhead{$m_{\lambda}$} & 
\colhead{$A_{\lambda}$} &
\colhead{$F_{\nu}$} &
\colhead{$\nu L_{\nu}$} \\
\colhead{} &
\colhead{} &
\colhead{(AB mag)} & 
\colhead{(mag)} &
\colhead{($\mu$Jy)} &
\colhead{(erg s$^{-1}$)}
}
\startdata
{\it Swift}/XRT & 1.7~keV & - & - & $<0.0128$ & $<1.28 \times 10^{42}$ \\
SDSS & $u$ & $19.974 \pm 0.545$ & 3.075 & $37.1 \pm 24.2$ & $(7.73 \pm 0.50) \times 10^{42}$ \\
Pan-STARRS & $g$ & $18.462 \pm 0.038$ & 2.276 & $149.7 \pm 5.3$ & $(2.29 \pm 0.08) \times 10^{43}$ \\
Pan-STARRS & $r$ & $17.904 \pm 0.029$ & 1.637 & $250.2 \pm 6.8$ & $(2.99 \pm 0.08) \times 10^{43}$ \\
Pan-STARRS & $i$ & $17.591 \pm 0.034$ & 1.186 & $333.9 \pm 10.6$ & $(3.27 \pm 0.10) \times 10^{43}$ \\
Pan-STARRS & $z$ & $17.419 \pm 0.028$ & 0.920 & $391.2 \pm 10.2$ & $(3.33 \pm 0.09) \times 10^{43}$ \\
Pan-STARRS & $y$ & $17.385 \pm 0.059$ & 0.759 & $403.7 \pm 22.5$ & $(3.09 \pm 0.17) \times 10^{43}$ \\
2MASS & $J$ & $16.964 \pm 0.119$ & 0.480 & $594.5 \pm 68.9$ & $(3.50 \pm 0.41) \times 10^{43}$ \\
2MASS & $H$ & $16.787 \pm 0.123$ & 0.283 & $700.2 \pm 83.4$ & $(3.12 \pm 0.37) \times 10^{43}$ \\
2MASS & $K$ & $16.707 \pm 0.115$ & 0.171 & $753.7 \pm 84.2$ & $(2.58 \pm 0.28) \times 10^{43}$ \\
WISE & $W1$ & $17.127 \pm 0.041$ & 0.075 & $511.9 \pm 19.7$ & $(1.12 \pm 0.04) \times 10^{43}$ \\
WISE & $W2$ & $17.443 \pm 0.048$ & 0.044 & $382.7 \pm 17.3$ & $(6.10 \pm 0.27) \times 10^{42}$ \\
WISE & $W3$ & $14.998 \pm 0.061$ & 0.013 & $3637.5 \pm 210.2$ & $(2.22 \pm 0.13) \times 10^{43}$ \\
WISE & $W4$ & $14.587 \pm 0.256$ & 0.007 & $5311.3 \pm 1412.3$ & $(1.76 \pm 0.47) \times 10^{43}$  \\
uGMRT & 650~MHz & - & - & $700 \pm 100$ & $(1.12 \pm 0.16) \times 10^{38}$ \\
VLA & 1.4~GHz & - & - & $<510$ & $\lesssim 1.76 \times 10^{38}$ \\
VLA & 3~GHz & - & - & $340 \pm 30$ & $(2.51 \pm 0.22) \times 10^{38}$ \\
VLA & 6~GHz & - & - & $221 \pm 21$ & $(3.26 \pm 0.31) \times 10^{38}$ \\
VLA & 9~GHz & - & - & $150 \pm 10$ & $(3.32 \pm 0.22) \times 10^{38}$ \\
\enddata
\tablecomments{Magnitudes $m_{\lambda}$ are corrected for Galactic extinction, $A_{\lambda}$, in the direction of the burst \citep{fm07}. Data are from the {\it Swift} archive (Program~1619112), SDSS~DR12 \citep{SDSS2020}, Pan-STARRS DR1 \citep{cmm+16}, 2MASS \citep{2MASS}, and AllWISE \citep{WISE,AllWISE}. uGMRT and VLA measurements are from \citet{atel14529}, \citet{atel14549}, and \citet{rll+21}. Upper limits correspond to the estimated $3\sigma$ confidence.}
\end{deluxetable*}

\subsection{{\it Swift}/XRT Observations}
\label{sec:xray}

The X-ray Telescope (XRT) on board the Neil Gehrels {\it Swift} Observatory ({\it Swift}; \citealt{ggg+04}) observed the location of \frb\ in a series of two observations: ObsID 00014258001, starting on 2021 April 6 at 19:48:08 UTC, and ObsID 00014258002, starting on 2021 April 7 at 06:30:37 UTC, under Program 1619112 (PI: L.~Piro) for 4.92~ksec each. Previous analyses of parts of these data yielded X-ray upper luminosity limits of $L_X \lesssim 2.2-4.4 \times 10^{42}$~erg~s$^{-1}$ (reported in \citealt{atel14523} and \citealt{atel14525}).

We retrieve the aforementioned {\it Swift}/XRT data from the \texttt{HEASARC} archive. We produce new event files for both observations using the \texttt{xrtpipeline} tool (\texttt{HEASoft} software, v.6.26; \citealt{Blackburn1999,NASA2014}) and \texttt{caldb} files (v.20200724). We stack the observations utilizing the \texttt{XSELECT} tool and obtain a total exposure time of 9.83~ksec. Within a source aperture of $20''$ in radius centered on the optical host centroid, we detect two counts in the 0.3-10.0~keV range. Using the Poisson single-sided upper limits introduced in \citet{geh86}, we calculate a $3\sigma$ upper limit of $<1.11 \times 10^{-3}$~cts~s$^{-1}$. To convert to flux, we employ the standard relation from \citet{wat11} between $A_V$ and $N_{\rm H,int}$, the hydrogen column density intrinsic to the host galaxy, using $A_V=1.5$~mag (Section~\ref{sec:prospector}) to obtain $N_{\rm H,int} = 3.3\times 10^{21}$~cm$^{-2}$. We also assume a standard power-law spectrum with a photon index $\Gamma_X=2$ (for consistency with \citealt{atel14523,atel14525} and other synchrotron sources \citealt{Ishibashi2010}) and the Galactic contribution to the hydrogen column density in the direction of the FRB, $N_{\rm H,MW}=4.49 \times 10^{21}$~cm$^{-2}$ \citep{wsb+13}, we calculate the unabsorbed X-ray flux to be $F_X\lesssim 1.09 \times 10^{-13}$~erg~s$^{-1}$~cm$^{-2}$ (0.3-10.0~keV). This translates to a $3\sigma$ upper limit on the X-ray luminosity of $\nu L_{X}\lesssim 1.3 \times 10^{42}$~erg~s$^{-1}$ at a central energy of 1.7~keV (Table~\ref{tab:hostphot}). This result is in agreement with the previous analyses reported in \citet{atel14523} and \citet{atel14525}.

\section{The Host Galaxy Properties of \frb}
\label{sec:host}

\subsection{Optical and Near-infrared Properties: A Dusty, Star-forming Galaxy}

Here, we explore the colors and emission line diagnostics of the host of \frb\ in the context of degree of global star formation. We first remark on the degree of contribution of the $z=0.553$ background galaxy to the existing host photometry in Table~\ref{tab:hostphot}. If we assume that the background galaxy has a luminosity of $L^{*}$, this corresponds to an apparent brightness of $r \approx 21.5$~mag, calculated from the galaxy luminosity function in the appropriate rest-frame band \citep{wfk+06}. Given the foreground host galaxy brightness of $r\approx 17.87$~mag, the background galaxy would contribute only $3\%$ to the total flux, or $0.04$~mag. Thus, we do not expect any background galaxy to contribute significantly to the flux and affect our subsequent conclusions.

At the most basic level, we first compare the colors of the host of \frb\ to characterize its nature in the context of other galaxies. The rest-frame $U-V$ and $V-J$ colors are often used to separate star-forming (SF) from quiescent galaxies on a $UVJ$ diagram \citep{wqf+09}. For the host of \frb\ at $z=0.0979$, the rest-frame $U-V$ and $V-J$ colors roughly correspond to $u-g \approx 1.6$~mag and $g-J\approx 1.4$~mag. In the context of the $UVJ$ diagram, the host of \frb\ exhibits redder $U-V$ colors than unobscured SF galaxies but is consistent with the colors of dusty SF galaxies \citep{flp+14,ffk+18}. Turning to the infrared, WISE colors may also be used to distinguish various star-forming galaxy types (e.g., normal star-forming, luminous infrared galaxies (LIRGs), and starbursts) from those which host quasars or obscured AGN \citep{WISE}. Using the extinction-corrected WISE colors (in the native Vega magnitude system), we find $W1-W2 \approx 0.31$~mag and $W2-W3 \approx 4.17$~mag. This corresponds to the parameter space occupied by LIRGs and starburst galaxies \citep{WISE}. Based on the WISE colors, we also note that the galaxy is unlikely to host a strong AGN, since the majority of such hosts have red ($W1-W2>0.6$~mag) colors \citep{Wu2012}.

We additionally explore the location of the host of \frb\ on the Baldwin-Phillips-Terlevich (BPT) diagram \citep{bpt81}, which distinguishes SF galaxies from low-ionization nuclear emission-line region (LINER) galaxies and those which host AGN according to their ratios of line fluxes, log([\nii]$\lambda$6583/H$\alpha$) and log([\oiii]$\lambda5007$/H$\beta$). To determine the line fluxes, $F_{\lambda}$, we use the {\tt Prospector} code (see Section~\ref{sec:prospector}; \citealt{Leja_2017,bdj21}) to fit a Gaussian line spread function with variable amplitude, width, and a small perturbation allowed around the measured line center for each generated model spectrum. This Gaussian sits on top of the model stellar spectrum, naturally including the effect of the underlying stellar absorption. Formally, the amplitude of the Gaussian is a compromise between the expectation value based on the CLOUDY photoionization model built into \texttt{Python-fsps} (Flexible Stellar Population Synthesis; \citealt{FSPS_2009, FSPS_2010}) and the observed spectrum (see Appendix E of \citealt{bdj21}); in practice, for the strong lines measured in this spectrum, the likelihood is dominated by the data, and the reported luminosities represent the observed luminosities corrected for the underlying stellar absorption. We report the line fluxes in Table~\ref{tab:nebemline}. These line fluxes yield ratios of log([\nii]$\lambda$6583/H$\alpha$)\,=\,$-0.35$ and log([\oiii]$\lambda5007$/H$\beta$)\,=\,$-0.51$. Compared to the BPT relations, this places the host of \frb\ in the region occupied by SF galaxies \citep{Brinchmann2008}. Based on an initial sample of eight FRB hosts, \citet{hps+20} found that they overall exhibit harder ionization fields than expected from the normal galaxy population, pointing toward a preference for LINER or AGN origins. On the contrary, the emission line flux ratios and optical/infrared (IR) colors classify the host of \frb\ as a dusty SF galaxy with no clear sign of AGN or LINER emission.

\subsection{Nebular Emission: Star Formation Rate and Metallicity}

\begin{deluxetable}{cc}[t!]
\linespread{1.2}
\tablecaption{\frb\ Host Nebular Emission Line Fluxes \label{tab:nebemline}}
\tablecolumns{3}
\tablewidth{0pt}
\tablehead{
\colhead{Line} &
\colhead{$F_{\lambda}$} \\
\colhead{} &
\colhead{($10^{-15}$ erg s$^{-1}$ cm$^{-2}$)}
}
\startdata
H$\beta$ & $1.23^{+0.30}_{-0.23}$ \\
{[}\oiii{]} $\lambda 4959$ & $0.12^{+0.07}_{-0.03}$ \\
{[}\oiii{]}$\lambda 5007$ & $0.38^{+0.21}_{-0.11}$ \\
{[}\nii{]}$\lambda 6548$ & $0.85^{+0.20}_{-0.16}$ \\
H$\alpha$ & $5.69^{+1.49}_{-0.99}$ \\
{[}\nii{]}$\lambda 6583$ & $2.53^{+0.60}_{-0.47}$ \\
{[}\sii{]}$\lambda 6716$ & $1.08^{+0.25}_{-0.20}$ \\
{[}\sii{]}$\lambda 6731$ & $0.84^{+0.19}_{-0.15}$ \\
\enddata
\tablecomments{These values are derived from {\tt Prospector}, and are corrected for underlying stellar absorption and Galactic extinction in the direction of the host.}
\end{deluxetable}

Given the SF designation of the galaxy, we use the observed H$\alpha$ line flux to obtain the host global star-formation rate (SFR). We correct the line emission for stellar absorption, and correct the H$\alpha$ line flux for the Balmer absorption (from stellar population modeling; Section~\ref{sec:prospector}), F$_\lambda$(H$\alpha$)/F$_\lambda$(H$\beta$)$\approx 6.6$, which is significant compared to the theoretical value for Case~B recombination in the absence of dust extinction of $\approx 2.86$ \citep{ost89}. Applying the \citet{cab+00} extinction law for actively star-forming galaxies yields a dust extinction of $A_V = 1.66 \pm 0.33$~mag and $A(H_\alpha) = 1.37 \pm 0.33$~mag. We calculate a dust-corrected line luminosity of $L(H\alpha) = 4.95^{+1.30}_{-0.86} \times 10^{42}$~erg~s$^{-1}$ yielding SFR$=2.12^{+0.69}_{-0.28}~M_{\odot}$~yr$^{-1}$ \citep{ken98,mkt+06}, assuming a Chabrier initial mass function (IMF) \citep{Chabrier2003}.

We infer the gas-phase metallicity of the host galaxy based on the strong-line diagnostics ratio, O3N2\footnote{O3N2 = log[([\oiii]$\lambda 5007$/H$\beta$)/([\nii]$\lambda 6583$/H$\alpha$)].}, of \citet{Hirschauer2018}, which yields an oxygen abundance of $12+\log({\rm O/H}) = 9.03^{+0.15}_{-0.24}$. We chose this calibration as it minimizes the effects of dust-obscuration on the individual line fluxes while including the large set of available lines detected in the spectrum. Relative to the stellar mass of the galaxy (see Section~\ref{sec:prospector}), the oxygen abundance is consistent with the mass-metallicity relation of SF galaxies at similar redshifts \citep{Maiolino2008}. 

\begin{figure*}[!t]
\centering
\includegraphics[width=\textwidth]{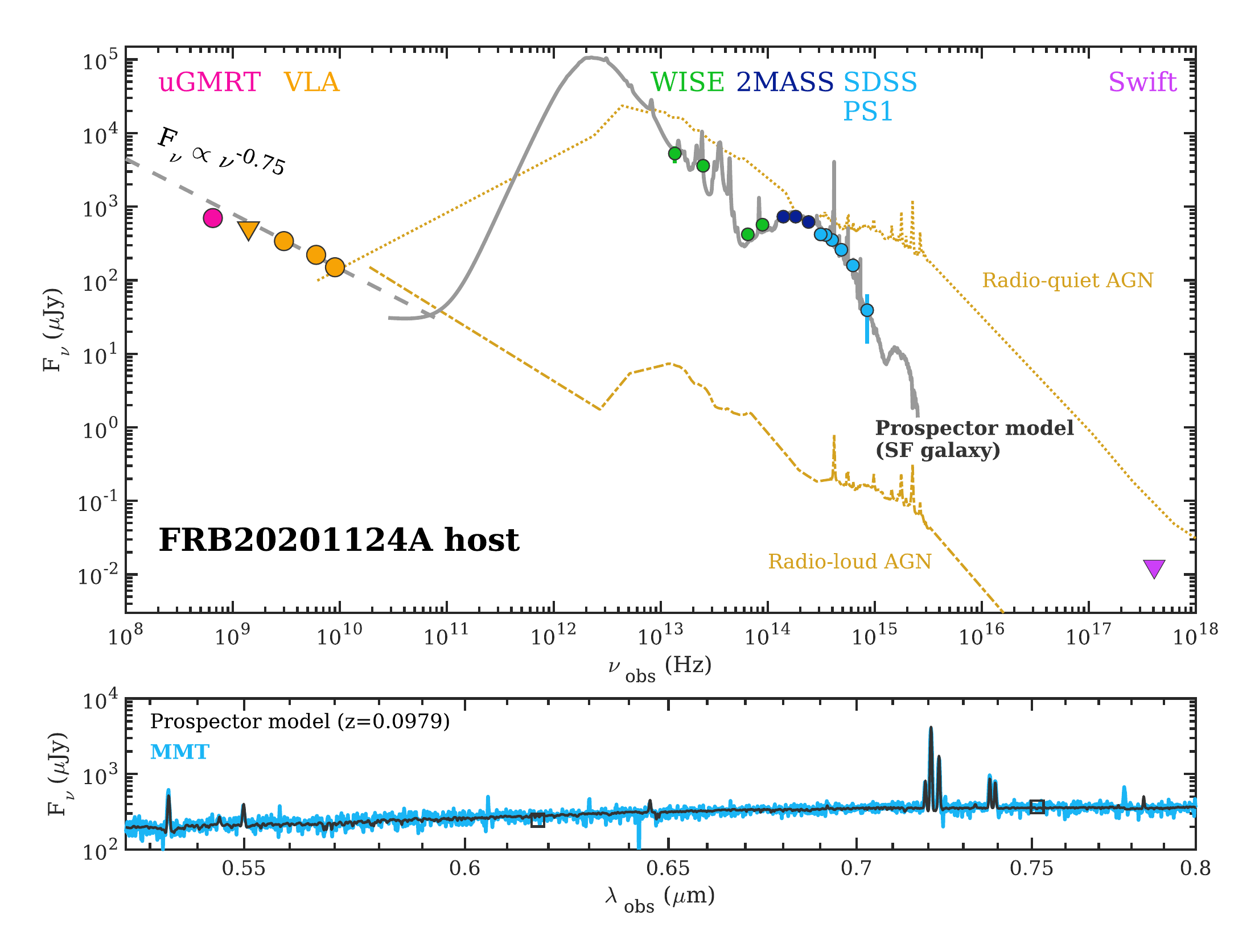}
\vspace{-0.4in}
\caption{{\it Top:} Broad-band SED of the host galaxy of \frb\ with data taken from uGMRT \citep{atel14529}, VLA \citep{atel14549,rll+21}, WISE, 2MASS, PS1, SDSS, and {\it Swift}/XRT (\citealt{atel14523,atel14525} and re-analyzed in this work). Circles denote detections and triangles denote $3\sigma$ upper limits. Also shown are the star-forming galaxy model from {\tt Prospector} (grey line) and the best-fit power-law to the radio data at $3~{\rm GHz} \lesssim \nu \lesssim 9$~GHz, representing a synchrotron component ($F_{\nu} \propto \nu^{-0.75}$; dashed line). The normal star-forming model provides a good fit to the data, whereas the radio-quiet and radio-loud AGN templates (tan dotted and dash-dotted lines; \citealt{sbw+11}), scaled to the radio fluxes, do not provide adequate matches to the broad-band data. {\it Bottom:} MMT spectrum (blue) versus wavelength, corrected for Galactic extinction, with the {\tt Prospector} model spectrum and photometry (black lines and squares) at $z=0.0979$, demonstrating agreement of the continuum shape and nebular emission lines.
\label{fig:bbsed}}
\end{figure*}

\subsection{Stellar Population Modeling}
\label{sec:prospector}

We determine the stellar population properties of the host galaxy of \frb\ with \texttt{Prospector}, a Python-based stellar population inference code \citep{Leja_2017,bdj21}. We jointly fit the observed photometry and spectroscopy using the nested sampling routine, \texttt{dynesty} \citep{Dynesty}, and build the model spectral energy distribution (SED) with \texttt{Python-fsps} \citep{FSPS_2009, FSPS_2010}. Our precise modeling assumptions and parameter definitions are described in Appendix~\ref{appendix}, and the priors on these properties are listed in Table~\ref{tab:priors}.

We jointly fit the observed photometry and spectrum of the host of \frb, both corrected for Milky Way extinction and weighted by the $1\sigma$ photometric uncertainties and error spectrum. We run multiple fits, employing a delayed-$\tau$ parametric SFH, and a non-parametric SFH. While the use of non-parametric SFHs has been demonstrated to represent a more physically realistic representation of galaxy histories \citep{Conroy2013SED,lcj+19}, such models are not yet broadly used in transient host galaxy literature. The choice of SFH also can result in known offsets of stellar population properties (cf., \citet{lcj+19}. Thus, including the derived properties of the parametric SFH fit here enables a direct comparison to other FRB hosts which have been modeled with similar assumptions, and to the {\tt Prospector} parametric SFH fitting results of \citet{rll+21} and \citet{pbt+21}.

The results of the parametric SFH and non-parametric SFH fits are listed in Table~\ref{tab:allprops}. In particular, we report the median values and bounds corresponding to 68\% credible intervals. The model representing the median of the non-parametric SFH posterior is plotted along with the observed photometry and spectroscopy in Figure~\ref{fig:bbsed}. In terms of stellar mass, dust content, and low stellar metallicity, we find similar values between our parametric and non-parametric SFH fits (Table~\ref{tab:allprops}). We further find a somewhat older median mass-weighted age of $t_m \approx 6.2$~Gyr in our non-parametric SFH fit. Comparing our parametric SFH results to the modeling results of \citet{rll+21} and \citet{pbt+21}, our results are broadly consistent although we report a mass-weighted age that is a bit older than implied by their fits.


Next, we investigate the cause of the low stellar metallicity. First, we remove the W3 and W4 fluxes from our fits, but still recover the same preference for low stellar metallicity. Next, we fix the stellar metallicity to the solar value and re-run the fit. This reveals that the continuum S/N in the spectroscopy is too low to put strong constraints on the metallicity. The low stellar metallicity is instead largely driven by the relatively blue near-IR colors (specifically the WISE W1-W2 color). \texttt{Prospector} is able to provide an adequate fit to the photometry assuming solar metallicity by dramatically lowering the emission from polycyclic aromatic hydrocarbon (PAH) molecules, thereby removing the strong 3.3$\mu$m PAH emission line. However, to explain the W3 and W4 fluxes without PAH emission would require an extraordinarily bright IR AGN with $\approx 58$ times the bolometric luminosity of the galaxy. We are unable to distinguish between the following three scenarios without more data: (a) the stellar metallicity is truly as low as measured, (b) there is an extremely bright IR AGN which has also destroyed all of the PAH molecules, or (c) the 3.3$\mu$m PAH  emission line alone is much dimmer in this galaxy than in galaxies with comparable amounts of PAH content. We note that removing the W3 and W4 fluxes entirely removes the need for a bright IR AGN.

\begin{deluxetable}{ccc}[t!]
\linespread{1.2}
\tablecaption{Derived Host Galaxy Properties of \frb\ \label{tab:allprops}}
\tablecolumns{3}
\tablewidth{0pt}
\tablehead{
\colhead{Property} &
\colhead{Value} & 
\colhead{Units}
}
\startdata
$z$ & $0.0979 \pm 0.0001$ \\
RA(J2000) & \ra{5}{08}{03.477}  \\
Dec(J2000) & $+$\dec{26}{03}{37.93} \\
$\delta R$ (VLBI) & $0.71 \pm 0.03$ & $''$ \\
$\delta R$ (VLBI) & $1.29 \pm 0.05$ & kpc \\
SFR (H$\alpha$) & $2.1^{+0.7}_{-0.3}$ & $M_{\odot}$~yr$^{-1}$ \\
SFR ($3-1100\,\mu m$) & $4.0^{+0.9}_{-0.5}$ & $M_{\odot}$~yr$^{-1}$ \\
SFR (radio)$^\dagger$ & $2.2-5.9$ & $M_{\odot}$~yr$^{-1}$ \\
12$+$log(O/H) & $9.03^{+0.15}_{-0.24}$ \\
log($L_B/L_{\odot}$) & 10.10 \\
\hline
\multicolumn{3}{c}{\emph{{\tt Prospector} Parametric Fit}} \\
\hline
log(M$_*$/M$_{\odot}$) & $10.23 ^{+0.02}_{-0.03}$ \\
$t_m$ & $4.97^{+0.36}_{-0.52}$ & Gyr \\
$t_{\rm max}$ & $11.00^{+0.92}_{-1.35}$ & Gyr \\
$\tau$ & $5.62^{+0.86}_{-1.02}$ & Gyr \\
log(Z$_*$/Z$_{\odot}$) & $-0.87 \pm 0.06$ \\
log(Z$_{\rm gas}$/Z$_{\odot}$) & $0.13 \pm 0.01$ \\
$A_{\rm V, old}$ & $1.02 \pm 0.04$ & mag \\
SFR$_{\rm SED}$ & $2.43 \pm 0.13$ & $M_{\odot}$~yr$^{-1}$ \\
\hline
\multicolumn{3}{c}{\emph{{\tt Prospector} Non-Parametric Fit}} \\
\hline
log(M$_*$/M$_{\odot}$) & $10.28 \pm 0.05$ \\
$t_m$ & $6.16^{+0.75}_{-0.80}$ & Gyr \\
log(Z$_*$/Z$_{\odot}$) & $-0.58 \pm 0.22$ \\
log(Z$_{\rm gas}$/Z$_{\odot}$) & $0.17^{+0.29}_{-0.19}$ \\
$A_{V, {\rm young}}$ & $0.91^{+0.44}_{-0.39}$ & mag \\
$A_{V, {\rm old}}$ & $0.81^{+0.20}_{-0.16}$ & mag \\
SFR$_{\rm SED}^{\ddagger}$ & $1.50^{+0.52}_{-0.35}$ & $M_{\odot}$~yr$^{-1}$ \\
sSFR(100~Myr) & $1.38^{+0.75}_{-0.49}$ & $10^{-10}$~yr$^{-1}$ 
\enddata
\tablecomments{Properties of \frb\ and its host galaxy determined in this work. The {\tt Prospector} parametric fit is characterized by a MW extinction law and delayed-$\tau$ SFH. Quoted upper limits correspond to the $99.7\%$ quantile. The {\tt Prospector} property definitions are described in Section~\ref{sec:prospector} \\
$\dagger$ Range set by frequency and calibration method \citep{yc02,mcs+11}. \\
$\ddagger$ Corresponds to ongoing SFR in most recent age bin.}
\end{deluxetable}

\subsection{Star Formation and Mass Assembly Histories}

\begin{figure*}[!t]
\centering
\includegraphics[width=0.49\textwidth]{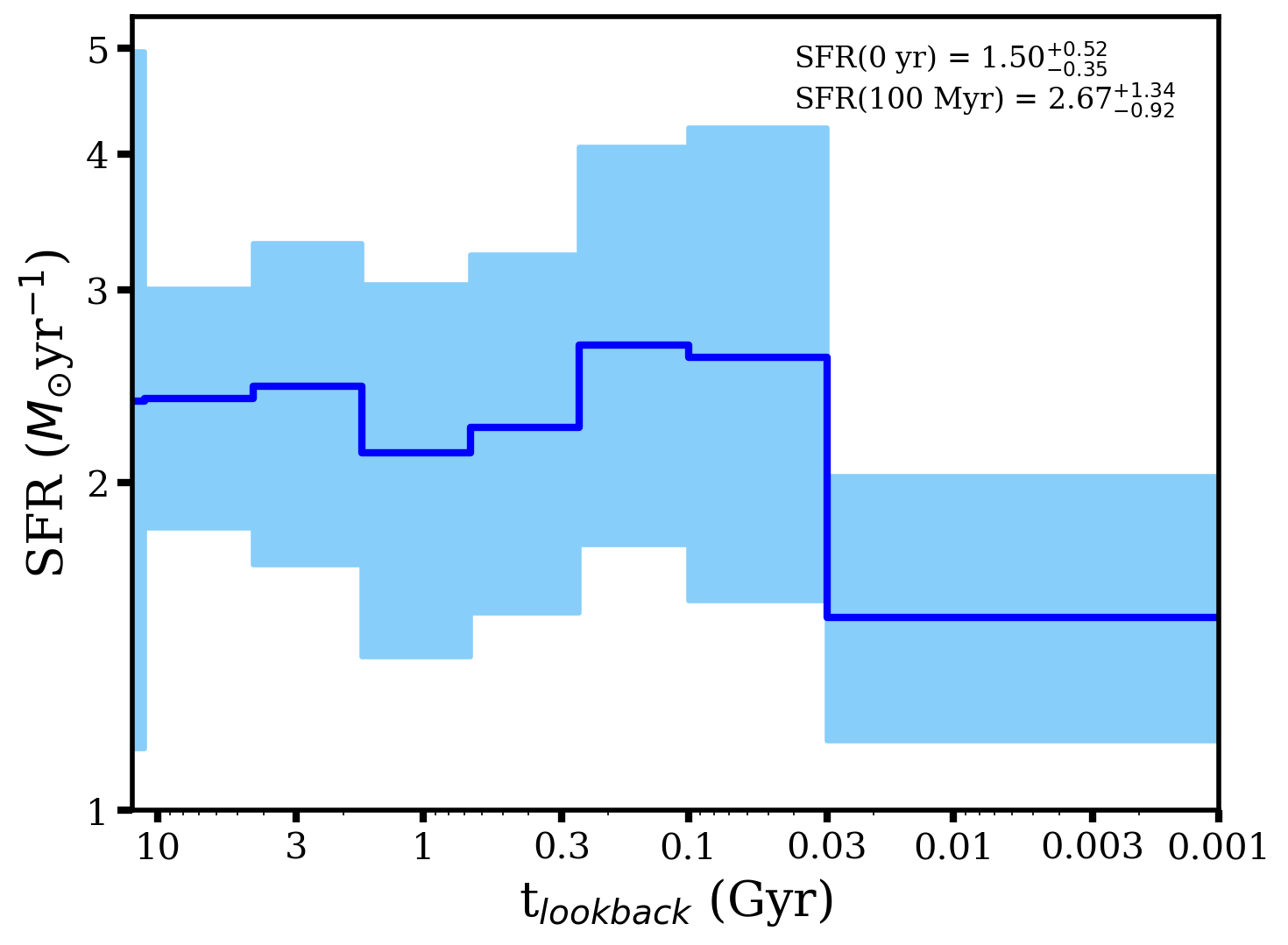}
\includegraphics[width=0.49\textwidth]{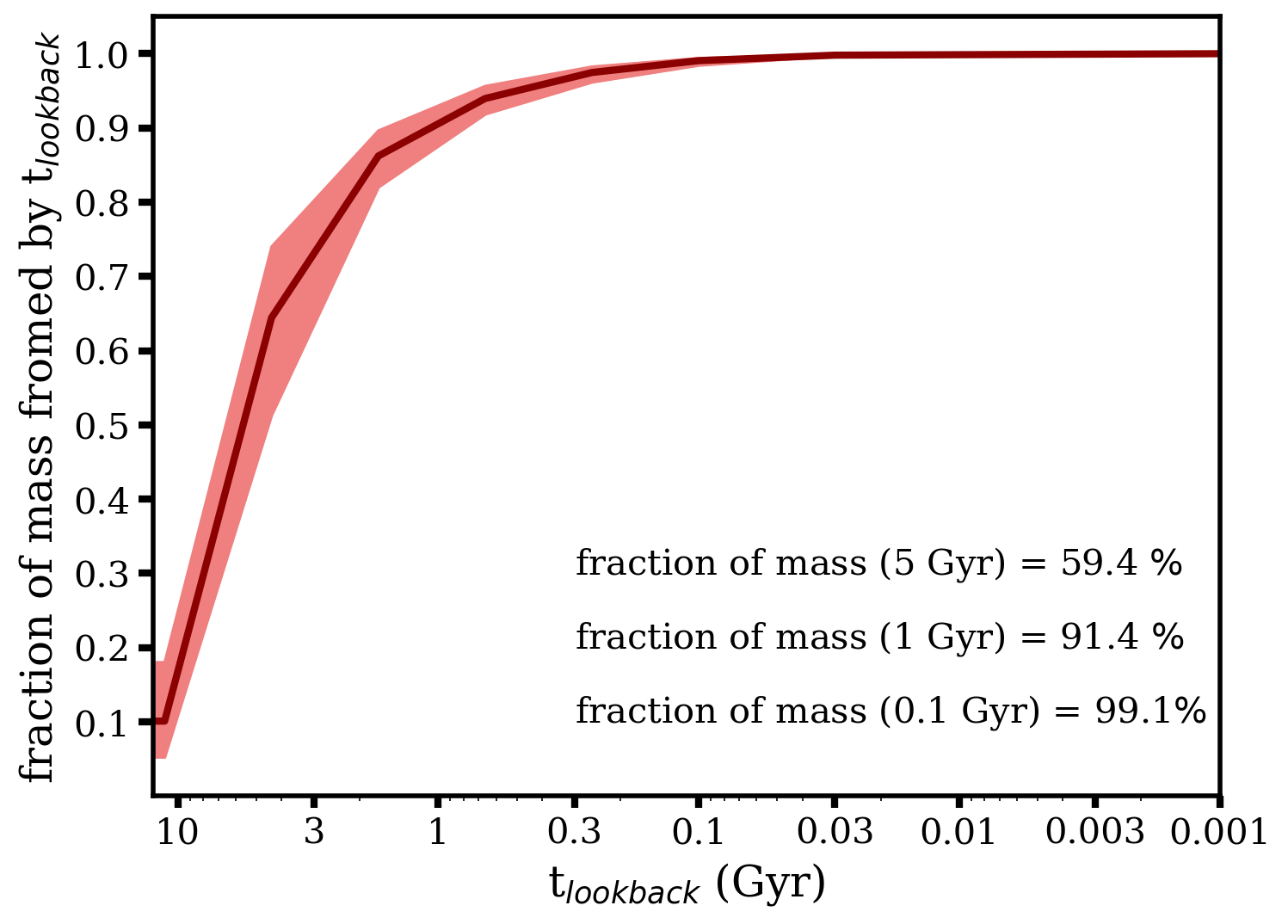}
\caption{{\it Left:} Derived SFR versus lookback time ($t_{\rm lookback}$) of the host of \frb\ for each of the eight time bins. The blue line represents the median, and the shaded region represents the 68\% quantile of each bin. The SFR was fairly constant for most of the host's history, with a notable absence of any star-burst. {\it Right:} Mass assembly history versus lookback time of the host of \frb\ (where $t_{\rm lookback}=0$ corresponds to the redshift of the host). The shaded region represents the 68\% quantile. By $t_{\rm lookback}=1$~Gyr (100~Myr), $\approx 91.4\%$ ($\approx 99.1\%$) of the galaxy's total mass was already formed. The history implies longer delay times for progenitors which trace stellar mass.
\label{fig:sfh}}
\end{figure*}

Using the results of the non-parametric SFH modeling, we are able to construct the star formation and mass assembly histories. The SFR as a function of lookback time (where $t_{\rm lookback}$ is defined in the frame of the FRB host) is displayed in Figure~\ref{fig:sfh}. We find that the SFR was roughly constant for most of the galaxy's history, at a rate of $\approx 2$ times the present-day value, before decreasing to $\approx 1.5\,M_{\odot}$~yr$^{-1}$ at $t_{\rm lookback}\approx 30$~Myr.

We also construct the cumulative mass assembly history of the host by calculating the fraction of mass using $N=500$ posterior samples of M$^*$(t) (i.e., the current mass in stars and stellar remnants as a function of creation time) from the {\tt Prospector} non-parametric fit\footnote{A sample-by-sample approach is necessary due to the strong bin-to-bin correlation in the mass formed in each time bin.}. For each posterior sample, the cumulative M$^*$(t) is constructed, and then this vector is divided by the total mass formed in that posterior sample. The reported values are the posterior median cumulative M$^*$(t) formed in each bin, and the uncertainties are taken as the 16\% and 84\% values of the posterior. The result is shown in Figure~\ref{fig:sfh}. A large majority of the stellar mass, $\approx 91.4\%$, was formed by $t_{\rm lookback}=1$~Gyr. By $t_{\rm lookback}=100$~Myr, 99.1\% of the host mass was formed by this time. Thus, despite the slight increase in the median absolute SFR at $t_{\rm lookback} \approx 30-300$~Myr, the peak was not prolonged or pronounced enough to contribute significantly to the total mass budget of the galaxy. In fact, within the 68\% confidence interval, the SFH is fairly constant until $t_{\rm lookback} \approx 30$~Myr.

We further find evidence for a hot dust emission component with $f_{\rm AGN} \approx 0.04$, corresponding to a mid-IR {\it luminosity} contribution of a potential AGN to the total mid-IR galaxy luminosity of $\approx 20\%$. This model parameter is typically interpreted as dust emission heated by a central AGN. However, we caution that there are other potential origins of dust heated to high temperatures, including extreme star formation \citep{ljc+18}, or higher centralized stellar densities \citep{gks+12}. Thus, while we cannot rule out the presence of an AGN completely, other multi-wavelength AGN diagnostics do not support an AGN as a strong contributor to the host of the \frb.

\subsection{Physical Origin of the Persistent Radio Emission}

We now address the origin of the persistent radio emission reported by uGMRT and the VLA \citep{atel14529,atel14549}. 
The lack of a counterpart in the milliarcsecond resolution observations reported in 1.374~GHz Very Long Baseline Interferometry (VLBI) observations \citep{atel14603,rll+21}, together with the fact that the persistent emission is unresolved to the VLA at 6~GHz with a beam size of $\approx 11''$ and an S/N$\approx 15$ \citep{rll+21}, constrain its size to the range $0.2-3''$ (or $\approx 0.4-5.5$~kpc at $z=0.0979$; see Section~\ref{sec:redshift}).
As it pertains to FRBs, there are three primary sources of persistent radio emission to consider: a compact radio source akin to that detected in association with FRB\,20121102A \citep{clw+17}, emission driven by disks, winds or jets from an AGN \citep{pbl+19}, or star formation in the galaxy \citep{con92}. The non-detection of persistent radio emission in high-resolution VLBI observations \citep{atel14603} immediately rules out a compact source (associated with either plerion-type emission like that seen in FRB\,20121102A or a core-dominated AGN) and instead points to an origin from a more extended spatial scale. The source is detected at 650~MHz (uGMRT; \citealt{atel14529}), 3~GHz, 6~GHz and 9~GHz (VLA; \citealt{atel14549,rll+21}), while there is an upper limit on the 1.4~GHz emission \citep{rll+21}. Emission was also reported at 22~GHz, although this emission is potentially co-located with the FRB position \citep{pbt+21}. The reported 650~MHz to 9~GHz fluxes, luminosities, and upper limits from these sources are compiled in Table~\ref{tab:hostphot}. The luminosity of the source is $\nu L_\nu\approx (2.5-3.3) \times 10^{38}$~erg~s$^{-1}$. Although the centroids of the reported uGMRT and VLA persistent radio emission are offset from each other, they are formally consistent within their quoted uncertainties (Figure~\ref{fig:host_image}). Thus, in what follows, we are treating the uGMRT and VLA emission as if they are originating from the same physical process, and the same general source.

We use $\chi^2$-minimization to fit power-law models to the data ($F_{\nu} \propto \nu^{\beta}$), incorporating the $1\sigma$ uncertainties on the fluxes. We find $\beta_{\rm {650MHz}\rightarrow{\rm 3GHz}}=-0.47 \pm 0.11$, with a steepening to $\beta_{\rm {3GHz}\rightarrow{\rm 9GHz}}=-0.75 \pm 0.13$. Fitting a broken power law model to all four data points, using the break frequency and smoothness as additional free parameters, we find it is only possible to constrain the location of the steepening to $\approx 4.7-7.6$~GHz with the available data, with the unconstrained smoothness of the break driving the uncertainty. This is in good agreement with the results from \citet{rll+21}. As the PRS is reported to be unresolved in the low angular resolution VLA and GMRT observations, there is no need to account for the impact of a changing beam size on the flux density obtained from these observations.

We next consider an AGN origin for the radio emission, as the reported locations of the persistent emission are also formally consistent with the host centroid (Figure~\ref{fig:host_image}). The non-detection of X-ray emission with {\it Swift}/XRT to an unabsorbed flux of $F_X \lesssim 1.1 \times 10^{-13}$~erg~s$^{-1}$~cm$^{-2}$ ($0.3-10$~keV; Section~\ref{sec:xray}), or $\nu L_\nu \lesssim 1.3 \times 10^{42}$~erg~s$^{-1}$, constrains ${\rm log}_{10}(L_R/L_X) \gtrsim -3.5$, placing any potential AGN in the ``radio-loud'' category \citep{tw03}. To demonstrate this, we adopt a radio-quiet AGN broad-band SED template \citep{sbw+11} and scale the model to the radio fluxes. Figure~\ref{fig:bbsed} demonstrates that this model both violates the X-ray upper limit and also over-predicts the IR and optical fluxes. On the other hand, the ${\rm log}_{10}(L_R/L_X)$ ratio is consistent with a radio-loud AGN, but a similarly-scaled template \citep{sbw+11} underestimates the optical/IR fluxes by three orders of magnitude (as these generally are prone to significant obscuration; Figure~\ref{fig:bbsed}). We further calculate the total IR luminosity ($L_{\rm TIR}$; $3-1100\,\mu$m) from the {\tt Prospector} model and find log($L_{\rm TIR}$/$L_{\odot}$)=$10.4^{+0.09}_{-0.06}$, which follows the canonical radio-TIR relations assuming pure star formation \citep{mds+21}. Thus, in support of the conclusions drawn from the emission line analysis, we consider it unlikely that an AGN is a dominant source of the observed persistent radio emission.

Finally, we consider whether the radio emission originates from recent star formation. Radio emission from galaxies is a combination of non-thermal, synchrotron emission from relativistic electrons accelerated by supernovae, with a canonical spectral index $\beta \approx -0.75$, and thermal, free-free emission with a flatter spectral index $\beta \approx -0.1$ \citep{con92}. At the frequencies observed for the FRB host (650~MHz to 9 GHz), the synchrotron component is expected to dominate, while there is an observed flattening at $\nu < 1$~GHz.

Given the constraints on the size of the emission region ($\approx 0.2-3''$ based on the non-detection on VLBI scales and it being unresolved with the VLA) and the fact that it is coincident with the nucleus of the galaxy, the PRS could be explained by diffuse circumnuclear emission, similar to the radio emission observed at the center of the Milky Way \citep{heywood+19} and other spiral galaxies \citep{ekers+74, condon+80}. Here, the radio emission is produced by a mixture of diffuse high energy electrons accelerated by a combination of supernovae (SNe), supernova remnants, ionized thermal emission, and highly magnetized filaments \citep{yhc+04}, all correlated with recent star-formation in the galactic center. In this scenario, the spectral flattening observed below 5\,GHz can also be attributed to free-free absorption by thermal emission embedded in this central region. In general, the radio emission could potentially be connected to the evidence of a hotter dust component from the optical and IR SED. We also note that \citet{pbt+21} suggests from high-frequency observations that the radio emission is co-located with the position of the FRB, although further high-resolution observations of the host are needed to precisely locate the region or regions of emission.

Using standard relations relating the radio continuum luminosity to the SFR, we derive SFR $\approx 2.2-5.9\,M_{\odot}$~yr$^{-1}$ \citep{yc02,mcs+11}, with the range depending on the SFR calibration used. From the total IR luminosity, we similarly derive SFR $\approx 4.00^{+0.93}_{-0.51}\,M_{\odot}$~yr$^{-1}$ \citep{hkj+11}. These derived values are slightly larger than the dust-corrected value inferred from the $H\alpha$ line emission\footnote{We note that the small differences in derived SFR can be attributed to a number of factors, including incomplete dust correction, contributions to the TIR luminosity the heating of dust, or the fact that the $H\alpha$-SFR conversion is systematically uncertain at the level of a factor of two (cf. \citealt{tss+19}).}.  Given this consistency, we consider recent star formation to be the dominant, if not the sole contributor to the observed persistent radio emission in the host of \frb.

\section{The Host Properties of \frb\ in Context: Implications for the Progenitor}
\label{sec:disc}

\frb\ now joins the small population of known repeating FRBs with identified hosts: FRB\,20121102A \citep{clw+17,tbc+17}, FRB\,20180916B \citep{mnh+20}, FRB\,20190711A \citep{hps+20,mpm+20}, FRB\,20200120E \citep{bgk+21}, FRB\,190520 \citep{lna+21}, and FRB\,20181030A \citep{bkm+21}. An additional fifth source of FRB-like emission, FRB\,200428, was traced back to a magnetar in our Galaxy \citep{CHIME-FRB200428}. Like the host of \frb, all of these galaxies are actively forming stars at low to moderate rates of $\approx 0.06-2\,M_{\odot}$~yr$^{-1}$ \citep{gpm+04,hps+20,lna+21}; the inferred SFR for the host of \frb\ is at the upper end of this range. The \frb\ host is the most massive when compared to the cosmological hosts beyond 5~Mpc \citep{hps+20} but is less massive than the Milky Way \citep{mcm11,ln15} and M81 \citep{dwb+08}, consistent with the stellar mass evolution of normal star-forming galaxies to low redshifts. Thus, in terms of specific SFR (i.e., SFR per unit stellar mass), the host of \frb\ is unremarkable compared to the host galaxies of other repeating FRBs, which span a wide range of $\approx (0.1-11) \times 10^{-10}$~yr$^{-1}$.

However, notable properties of the host in terms of the repeating FRB host population include its significant dust attenuation and potentially unusually low stellar metallicity of $\approx\!(0.1-0.3)Z_{\odot}$ (although we note that the gas-phase metallicity is higher at $\approx\!Z_{\odot}$). In terms of their emission line ratios, five other FRB hosts, including that of one other repeater (FRB\,20121102A), lie in the general plane of SF galaxies \citep{hps+20}. Beyond \frb, only FRBs\,20191001, and 20190608 have hosts with faint radio emission interpreted as star formation \citep{bbl+20,bsp+20}\footnote{We note that the candidate host of FRB\,20171020A also has faint radio SF, but its host association is less secure \citet{mem+18}}. However, unlike \frb, all of these FRBs are apparent non-repeaters.

At a projected physical offset of $\approx\!1.3$~kpc, \frb\ is also more proximal to its host center than most FRBs with known hosts, in the lower $\approx 10\%$ of the offset distribution \citep{hps+20,mfs+20}. From the available data, the FRB position does not appear to be coincident with any flux enhancements, such as a spiral arm, as has been seen for some of the other FRBs \citep{mfs+20}. However, higher-resolution imaging is needed to assess the presence of any underlying sub-structure.
 
Compared to the general galaxy population, the host of \frb\ follows the standard relations for normal, star-forming galaxies at $z\sim 0.1$. The host lies on the star-forming main sequence of galaxies \citep{wvb+12}, demonstrating that it is forming stars at a similar rate to other galaxies of the same stellar mass. Its gas-phase metallicity and stellar mass also follow the mass-metallicity relation \citep{thk+04,Maiolino2008}. While the host of \frb\ exhibits significant dust content, this is not necessarily unexpected for star-forming galaxies in its mass range \citep{wvb+12}. The host SED additionally supports a hot dust component contributing at the level of 20\% to the mid-IR luminosity, which cannot easily be attributed to a strong AGN given the broadband data. However, the dust heating could be the result of higher stellar densities (as seen in M31; \citealt{gks+12,vbt+17}), and potentially related to the persisent radio emission on $\lesssim 5$~kpc size scales, which we conclude to be due to recent circumnuclear star formation. 

The properties and history of \frb's stellar population provide direct points of comparison to expectations for various progenitors. We first consider the production of (millisecond) magnetars through ``ultra-prompt'' channels, such as Type I super-luminous SNe (SLSNe) or long-duration gamma-ray bursts (LGRBs) (e.g., \citealt{mmk+15,mbm17,ebm+19}). Such magnetars may naturally acquire their magnetic fields from their parent stars or via a convective dynamo shortly after collapse (e.g., \citealt{dt92,rgj+20}). The existence of the persistent radio source coincident with FRB\,20121102A \citep{clw+17} has been interpreted as emission from a young magnetar nebula or low-luminosity AGN \citep{mbm17,mm18,km17,msh+18,ebm+19}. The FRB source is also located in a star-forming knot in its dwarf host galaxy, similar to the properties of LGRBs and Type I SLSNe \citep{bta+17,tbc+17}; taken together, these properties make ultra-prompt channels a compelling progenitor scenario to consider for repeating FRBs.

However, the hosts of LGRBs and SLSNe have median stellar masses and luminosities an order of magnitude lower than the host of \frb\ \citep{sgl09,vsj+15,pqy+16,sys+20}; indeed it is rare for a SLSN or LGRB host to have an inferred stellar mass of $>10^{10}\,M_{\odot}$ when considering similar redshifts. In addition, unlike what is observed in the host of \frb, the host galaxies of LGRBs and SLSNe each exhibit a strong connection with enhanced global and local star-formation, or starburst activity \citep{lbk+10,lcb+14,pph+15,hth+18}. In contrast, \frb\ host's star formation rate is remarkably constant over its $>10$~Gyr history, with a recent decrease only in the last $\approx 30$~Myr by a modest factor of $\approx 2$. In the context of an ultra-prompt progenitor channel, we do not find clear evidence of any past or recent star-burst activity that could be connected to the birth of a very young ($\lesssim$~few Myr) progenitor.

Thus, the host properties of \frb, together with the lack of a compact radio remnant, contribute to the growing body of circumstantial evidence that such ``ultra-prompt'' channels are not dominant in the production of observed FRBs, and even repeating FRBs \citep{bsp+20,hps+20,lz20,sph+20}. On the other hand, as a fairly typical star-forming galaxy, the host properties of \frb\ are consistent with those of core-collapse SNe (CCSNe; e.g., \citealt{kk12,gar+16,sys+20}). Magnetars produced in such ``prompt'' channels have been argued to be a dominant channel for the extragalactic population of FRBs (e.g., \citealt{brd21}, see also \citealt{clg+21}). The delay times of CCSNe range from $\approx$few to 200~Myr, with a median of $\approx 22$~Myr \citep{zdi+17,cag+21}, which could be accommodated given the fairly constant host SFH.

Given the moderate mass-weighted stellar population age of $\approx 5-6$~Gyr, the host of \frb\ can also accommodate progenitors that have longer delay times. Such ``delayed'' channels include the production of magnetars from compact object-related systems, including neutron star mergers, or the accretion-induced collapse (AIC) of a white dwarf (WD) to a neutron star (NS) \citep{mbm+19}.  Such systems could have easily been formed earlier in the history of the host of \frb, only to produce a magnetar (and resulting observable FRB activity) fairly recently. Further still, dynamical formation channels of compact object systems in globular clusters can have long delay times due to the presence of binaries or black holes, which inhibits mass segregation and interactions \citep{kyc+20}. This was proposed as one of the explanations for the extremely old $\approx 9.1$~Gyr stellar population of FRB\,20200120E \citep{kmn+21}.

The broad family of delayed channels is flexible in producing magnetars across a wide range of timescales, thus resulting in diverse host galaxy demographics \citep{nwb+17,mbm+19}. However, for each of these channels which (in part) trace host stellar mass, some fraction of FRBs should {\it also} appear in quiescent galaxies. In general, for progenitors which trace stellar mass alone, the FRB host galaxy demographics should reflect the Universe's stellar mass budget, which is roughly 1:1 in quiescent and SF galaxies at $z\lesssim 0.5$ \citep{bmk+03,isl+10}. For instance, a substantial fraction of the host galaxies of short-duration gamma-ray bursts (derived from NS mergers) and Type Ia SNe (derived from WD progenitors) are quiescent \citep{slp+06,fbb+17}. Although there are now only seven known repeating FRBs with identified hosts to date, the lack of any quiescent host galaxy among this host population to date indicates that delayed channels are likely not responsible for all repeating FRBs, or that there is some uncharacterized bias against detecting repeating FRBs in quiescent, old stellar populations. However, the fact that a few repeating FRB hosts fall below the star-forming main sequence \citep{bha+21}, coupled with the the old local environment of FRB\,20200120E also indicates that recent or ongoing SF are not playing a dominant role in the production of FRBs.

\section{Summary}
\label{sec:conc}

We have presented the ASKAP localization of the bright, repeating \frb\ source, as well as detailed observations and modeling of its host galaxy at $z=0.0979$. Compared to the host properties of the FRB population, we find that the host galaxy is modestly star-forming ($\approx 2-6\,M_{\odot}$~yr$^{-1}$), moderately massive ($\approx 2 \times 10^{10}\,M_{\odot}$), dusty ($\approx 1-1.5$~mag of attenuation at optical wavelengths), and has a substantial stellar population (mass-weighted) age ($\approx 5-6$~Gyr). We also find that the host has a somewhat low stellar metallicity of $\approx (0.1-0.3)Z_{\odot}$. The properties of the host place it among the population of normal star-forming galaxies at $z\sim 0.1$ in terms of star formation, stellar mass, and gas-phase metallicity. The higher stellar mass of the host of \frb\ is commensurate with its elevated SFR, and the specific SFR is unremarkable compared to the repeating FRB host population. Modeling the star formation and mass assembly histories, there is no clear evidence for any star-burst activity, and most of the galaxy's stellar mass was built prior to 1~Gyr ago. We further find that the primary source of the reported persistent radio emission is recent star formation, as opposed to strong AGN activity.

\frb\ marks the fifth extragalactic repeating FRB with an identified host galaxy. Notably, {\it all} identified hosts of known repeating FRBs exhibit modest amounts of ongoing star formation. However, no repeating FRB yet is localized to a quiescent galaxy, and several repeating FRB hosts are forming stars at a lower rate than field galaxies of the same stellar mass, indicating that ongoing SF is also not playing a dominant role in the FRB rate. This study demonstrates the advantage of using complementary tools -- host galaxy demographics, local environments, and now star formation histories -- to decipher the progenitors of FRBs. In particular, building the histories of these host galaxies provides an important view beyond the ``present-day'' snapshots afforded by more traditional stellar population modeling. Moreover, given the fairly sudden onset of activity of the \frb\ source, as well as its brightness, it will be particularly insightful to connect the context clues learned from environments to the diversity of observed repeating FRB behaviors.

\section*{Acknowledgements}
We thank MMT Observatory staff Benjamin Weiner and Skyler Self for rapidly scheduling and executing key obserevations. We thank Ron Ekers, Brian Metzger and Sergio Campana for clarifying discussions. The Fast and Fortunate for FRB Follow-up team acknowledges support by the National Science Foundation under grant Nos. AST-1911140 and AST-1910471. The Fong Group at Northwestern acknowledges support by the National Science Foundation under grant Nos. AST-1814782, AST-1909358 and CAREER grant No. AST-2047919.  RMS acknowledges support through ARC Future Fellowship FT190100155. ATD is the recipient of an Australian Research Council Future Fellowship (FT150100415). 
NT acknowledges support by FONDECYT grant 11191217.
This research was partially supported by the Australian Government through the Australian Research Council's Discovery Projects funding scheme (project DP210102103).

MMT Observatory access was supported by Northwestern University and the Center for Interdisciplinary Exploration and Research in Astrophysics (CIERA). Observations reported here were obtained at the MMT Observatory, a joint facility of the University of Arizona and the Smithsonian Institution. This work made use of data supplied by the UK Swift Science Data Centre at the University of Leicester. This research was supported in part through the computational resources and staff contributions provided for the Quest high performance computing facility at Northwestern University which is jointly supported by the Office of the Provost, the Office for Research, and Northwestern University Information Technology. The National Radio Astronomy Observatory is a facility of the National Science Foundation operated under cooperative agreement by Associated Universities, Inc. 
The Australian SKA Pathfinder is part of the Australia Telescope National Facility which is managed by CSIRO. Operation of ASKAP is funded by the Australian Government with support from the National Collaborative Research Infrastructure Strategy. ASKAP uses the resources of the Pawsey Supercomputing Centre. Establishment of ASKAP, the Murchison Radio-astronomy Observatory and the Pawsey Supercomputing Centre are initiatives of the Australian Government, with support from the Government of Western Australia and the Science and Industry Endowment Fund. We acknowledge the Wajarri Yamatji people as the traditional owners of the Observatory site.

\vspace{5mm}
\facilities{MMT (Binospec), ASKAP, VLA, Swift (XRT)}

\software{\texttt{HEASoft} software \citep[v.6.26;][]{Blackburn1999,NASA2014}, \texttt{Prospector} \citep{Leja_2017}, \texttt{Python-fsps} \citep{FSPS_2009, FSPS_2010}, \texttt{Dynesty} \citep{Dynesty}, PypeIt \citep{{PypeIt}}}

\appendix 
\restartappendixnumbering

\section{Prospector Stellar Population Modeling Description} \label{appendix}

\begin{deluxetable}{cc}[t!]
\linespread{1.2}
\tablecaption{Prior Ranges for {\tt Prospector} Fitting \label{tab:priors}}
\tablecolumns{2}
\tablewidth{0pt}
\tablehead{
\colhead{Property} &
\colhead{Range} 
}
\startdata
log(M$_F$/M$_{\odot}$) & [8, 12]  \\ 
log($r_i$) & [-100,100]$^{\ddagger}$ \\
$t_{\rm age}^{\dagger}$ (Gyr) & [0, 12.4] \\
$\tau^{\dagger}$ (Gyr) & [0.1, 10] \\
log(Z$_*$/Z$_{\odot}$) & [-2.0, 0.19] \\ 
log(Z$_{\rm gas}$/Z$_{\odot}$) & [-2, 0.5] \\
dust1/dust2 & [0, 1.5] \\ 
dust2 & [0, 4] \\ 
$f_{\rm AGN}$ & [$10^{-5}$, 3.0] \\
$\tau_{\rm AGN}$ & [5.0, 150.0] 
\enddata
\tablecomments{$\dagger$ For parametric SFH fits only. \\
$\ddagger$ For non-parametric SFH fits only.}
\end{deluxetable}

Here, we describe the details of our {\tt Prospector} stellar population modeling parameter definitions, priors, and assumptions. We initialize our stellar population models with a Chabrier IMF \citep{2003PASP..115..763C} and Milky Way attenuation law \citep{MilkyWay} and further impose a few priors on the fit. First, we employ two wavelength- and age-dependent dust screens to represent (a) the attenuation by dust of stellar birth clouds, which affects only young stars ({\tt dust1} in Prospector), and (b) the attenuation by dust of the diffuse ISM, which affects both young and old stars ({\tt dust2} in Prospector). We assume a transition time between our designation of ``young'' and ``old'' stars of $10^{7}$~yr \citep{Conroy2013SED} and impose a 2:1 ratio on the amount of dust attenuation between the younger and older stellar populations, as young stars in SF regions typically experience twice the amount of dust attenuation as older stars \citep{cab+20, pkb+14}. We additionally impose that the posteriors roughly adhere to the mass-metallicity ($M_*-Z$) relationship \citep{gcb+05} at the relevant redshift. We employ a sixth-order Chebyshev polynomial to fit the spectral continuum.

We fix the redshift to $z=0.0979$ and determine posteriors for all other free parameters, which include the total mass formed of stars from dust over the lifetime of the galaxy ($M_F$), the lookback time at which star formation commences in the frame of the FRB host ($t_{\rm max}$, also corresponding to the maximum age), the stellar and gas-phase metallicities ($Z_*$, $Z_{\rm gas}$), the $e$-folding time $\tau$ in the assumed delayed-$\tau$ star formation history (SFH), and dust attenuation. We further use the samples of $\tau$ and $M_F$ to calculate the posteriors of the present-day stellar mass ($M_*$), mass-weighted age ($t_m$), SFR, and $A_{V}$ using analytic conversions \citep{Leja2013,nfd+20}. For fits with mid-IR components, $f_{\rm AGN}$ represents the fraction of bolometric luminosity contribution from the AGN, and $\tau_{\rm AGN}$ is the optical depth of the dust torus. The assumed priors on these properties are listed in Table~\ref{tab:priors}.  To test how the results are affected by some of our input assumptions, we perform two additional fits with a delayed-$\tau$ SFH: one using the Calzetti attenuation law \citep{cab+00}, which is broadly applicable to SF galaxies, and one which includes a mid-IR AGN component. For a fit which includes the mid-IR AGN contribution, we find a fraction of AGN contribution to the total flux of $\approx 6\%$, and its inclusion has little effect on the other derived properties.

For the non-parametric SFH fit, we include several additional spectroscopic calibration parameters and a flexible dust attenuation law following the model in Section 4.3 and Table~2 of \citet{bdj21}. We describe the SFH with eight bins in lookback time, two spaced at $0-30$ Myr and $30-100$ Myr and the remaining six spaced evenly in logarithmic time, with the upper limit set to be the age of the universe at the observed redshift. These are characterized by the parameter $r_i$ in the fit, which represents the ratio of the SFR in temporal bin $i$ to that in the adjacent bin. A continuity prior is adopted for the SFH \citep{lcj+19}, which weights against strong changes in the SFR with time and prefers a constant SFR in the absence of data. Thus, any deviation from a flat star formation history in the results will be driven by the data and not the priors. In this fit, we also include a mid-IR AGN component. The priors are listed in Table~\ref{tab:priors}.

\bibliography{refs}

\end{document}